\documentclass[twocolumn]{article}

\usepackage[utf8]{inputenc}
\usepackage{siunitx}
\usepackage{array,url,kantlipsum}
\usepackage[pdftex]{graphicx}
\newcommand{\Mark}[1]{\textsuperscript{#1}}
\usepackage{hyperref}
\usepackage[super,sort&compress,comma]{natbib} 
\usepackage[left=1.7cm, right=1.7cm, top=1.785cm, bottom=2.0cm]{geometry}
\usepackage[format=plain,justification=justified,singlelinecheck=false,font={stretch=1.125,small,sf},labelfont=bf]{caption} 
\usepackage{balance}
\usepackage{textcomp}
\usepackage{xcolor}
\usepackage{lmodern}
\usepackage{gensymb}
\usepackage[version = 4]{mhchem}

\begin{document}
 \sffamily
\twocolumn[{%

\LARGE Nanodroplet-Confined Electroplating Enables Submicron Printing of Metals and Oxide Ceramics\\[1.em]

\large Mirco Nydegger\Mark{1,2}$^\dag$,
        Rebecca A. Gallivan\Mark{1,3,}$^\dag$$^{*}$,
        Arthur Barras\Mark{1},
        Henning Galinski\Mark{1},
        and Ralph Spolenak\Mark{1,}$^{*}$
       \\[1em]

\normalsize
 
\Mark{1}\small{Laboratory for Nanometallurgy, Department of Materials, ETH Zürich, Vladimir-Prelog-Weg 5, 8093 Zürich , Switzerland}\\ 

\Mark{2}\small{Present address: Department of Aeronautics and Astronautics, MIT, 77 Massachusetts Avenue, Cambridge, MA 02139, USA}\\ 

\Mark{3}\small{Present address: Thayer School of Engineering at Dartmouth College, 15 Thayer Drive, Hanover, NH 03755, USA}\\ 
 
$^\dag$These authors contributed equally\\
  
$^{*}$To whom correspondence should be addressed; e-mail: rebecca.gallivan@dartmouth.edu; ralph.spolenak@mat.ethz.ch\\[0.5em]

\textbf{The fabrication of functional micro- and nano-electronic devices requires the deposition of high-quality materials of different electronic material classes, such as conductors, semiconductors and insulators. To establish ultra-high-resolution additive manufacturing as a viable addition to existing fabrication methods requires the combinatorial additive deposition of different electronic material classes. However, current techniques do not provide such a capability. Here, we demonstrate that droplet confined electroplating, an ultra-high-resolution AM technique initially developed for metals as electrohydrodynamic redox printing (EHD-RP), allows not only the direct deposition of many metals, but also of metal-oxides. Particularly, we demonstrate that applying fundamental electrochemical principles in combination with on-the-fly switching of the deposited material allows for the direct co-deposition of metals, metal-hydroxides and -oxides. Our results exemplify the feasibility of leveraging simple water-based electrochemical concepts to produce intricate and multi-material structures at the nanoscale.}\\[1em]
 
}]

\section{Introduction}

\begin{figure*}[!ht]
 \centering
 \includegraphics{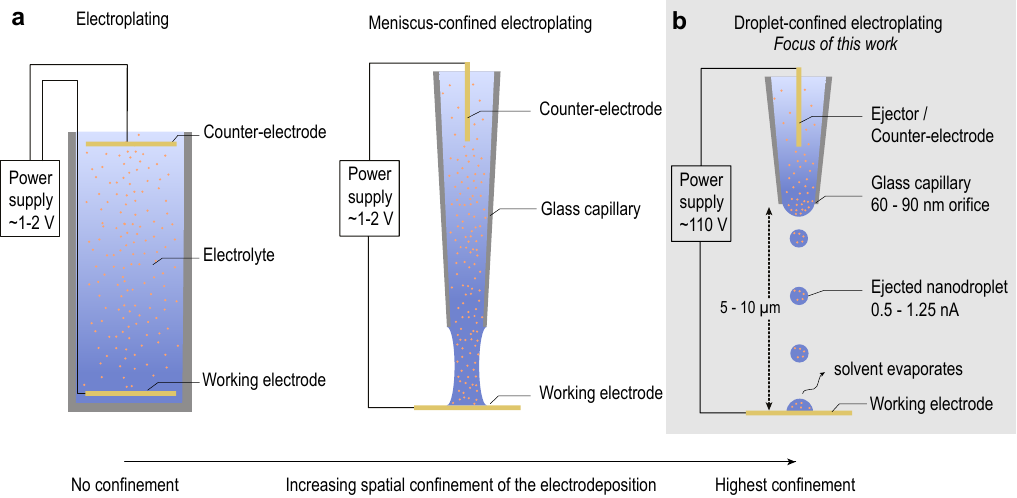}
 \caption{\textbf{The two-electrode setup: from standard electroplating to droplet-confined electroplating. (a)} Meniscus-confined electroplating confines the electrodeposition of metals into a small volume  by using a liquid bridge between on a solvent-filled capillary and a conductive substrate. This process is still very similar to normal electroplating, as both electrodes are connected through an electrolyte. \textbf{(b)} Droplet-confined electroplating is similarly based on a solvent-filled capillary in close proximity to a conductive substrate \cite{Reiser2019Multi-metalScale}. Applying a high electric potential between an electrode immersed within the capillary and the conductive substrate leads to an electrohydrodynamic ejection of an ion-containing droplet. This droplet then impacts on the substrate, the contained ions are reduced and the droplet evaporates. A repetition of this cycle leads to the formation of an out-of-plane metal nanostructure. Notably, no liquid connection is maintained between the electrolyte reservoir and the individual droplets \cite{Nydegger2024Droplet-ConfinedNanowires, Menetrey2024OnPrinting}.}
  \label{Technique}
\end{figure*}

Electroplating is a long-established and foundational technique in materials science that enables the controlled deposition of metals and compounds onto conductive substrates through electrochemical reduction\cite{2010ModernElectroplating}. The technique relies on an ionic exchange through a continuous electrolyte medium and is therefore traditionally performed in bulk electrolytic cells with well-defined anode–cathode configurations to deposit metal coatings. However, miniaturized and localized versions of this process, such as meniscus-confined electroplating \cite{Hengsteler2021BringingNanoscale, Hengsteler2022ElectrochemicalPerspective}, are emerging as powerful tools for additive manufacturing at the micro- and nanoscale. In this confined geometry, a liquid bridge is used to deliver the electrolyte to a targeted surface region, thus unlocking the free-form deposition of metallic structures (Figure ~\ref{Technique}a)~\cite{Hu2010Meniscus-confinedBonds}. The approach of spatially confining the electrolyte has been pushed even further through electrohydrodynamic redox printing (EHD-RP) where electroplating is confined within isolated droplets of down to 50~nm in diameter (Figure ~\ref{Technique}b) \cite{Menetrey2024OnPrinting}. The individual droplets, containing dissolved metal ions, are brought into contact with a conductive substrate by using electrohydrodynamic ejection from a nozzle\cite{Reiser2019Multi-metalScale, Menetrey2024NanodropletPrinting}. On the substrate, metal ions are deposited while the solvent evaporates. The result is a localized and transient electrochemical cell (Figure ~\ref{Technique}b) that evolves dynamically with the applied electric field and depends on the distance between the ejector and substrate\cite{Nydegger2024Droplet-ConfinedNanowires}. The ejection frequency of the individual droplets is estimated to be $\approx$ 2.5~MHz \cite{Menetrey2022TargetedPrinting} and the current is in the range of 0.5-1.25~nA (which relates to a current-density within the deposited structure of 1.5 \textendash 4 A cm\textsuperscript{-2} for a pillar with 200~nm diameter). The change from a meniscus-confined liquid bridge to confinement inside a droplet brings four major benefits: first, the forced mass transfer through a continuous stream of droplets allows for a higher deposition speed than a diffusion-limited process would\cite{Hirt2017AdditiveScale, Menetrey2022TargetedPrinting, Park2007High-resolutionPrinting, Liashenko2020}. Second, supplying droplets loaded with ions of different elements allows for a rapid change in the deposited metal\cite{Reiser2019Multi-metalScale}, granting access to deposit structures with a chemical architecture. Third, the confined environment unlocks unique pathways for material synthesis, including access to metastable phases and non-equilibrium structures \cite{Porenta2023Micron-scalePrinting}. Lastly, the deposition of individual droplets can be combined with an external source of electrons, such as an electron beam, to allow the deposition of metals also onto non-conductive substrates \cite{Nydegger2024DirectSize}.   

Importantly, in droplet-confined electroplating, the electrolyte on the working electrode is no longer part of a continuous bath connected to a distant counter-electrode \cite{Nydegger2024Droplet-ConfinedNanowires, Menetrey2024OnPrinting}. Instead, the spatial isolation introduces a fundamental change to the electrochemical environment: ionic exchange with a bulk reservoir is eliminated, and electrode reactions proceed under isolated conditions. As a result, traditional electroplating recipes presumably cannot be translated directly to the confined regime. This leads to a limited range of metals that have been reported so far (namely only Cu, Ag and Zn have been reported previously \cite{Reiser2019Multi-metalScale, Nydegger2022AdditiveStructures, Porenta2023Micron-scalePrinting}), but also to frequent observation of contaminants in the deposited structures \cite{Porenta2023Micron-scalePrinting}. Hence, the range of materials and a strategy to improve purity of the deposited structures are challenges that need to be resolved before droplet confined electroplating can be used to create functional structures.  

In the present study, we deposit a range of elements with standard electrode potentials between 1.83~V and -2.57~V. This systematic study enables us to derive a framework that specifically considers the role of speciation and chemistry of the ejected droplet for the deposition. Such a framework may guide future research towards selecting ideal concentrations and chemistries for different elements. We then demonstrate that the established framework allows for the deposition of materials with complex chemistry by utilizing the controlled co-deposition of salts. Lastly, we explore the deposition of composite structures \textendash potentially opening simple pathways to additive manufacturing of functional metal/metal-oxide structures at the nanoscale.

\section{Results}

\begin{figure*}
\centering
 \includegraphics[scale=1]{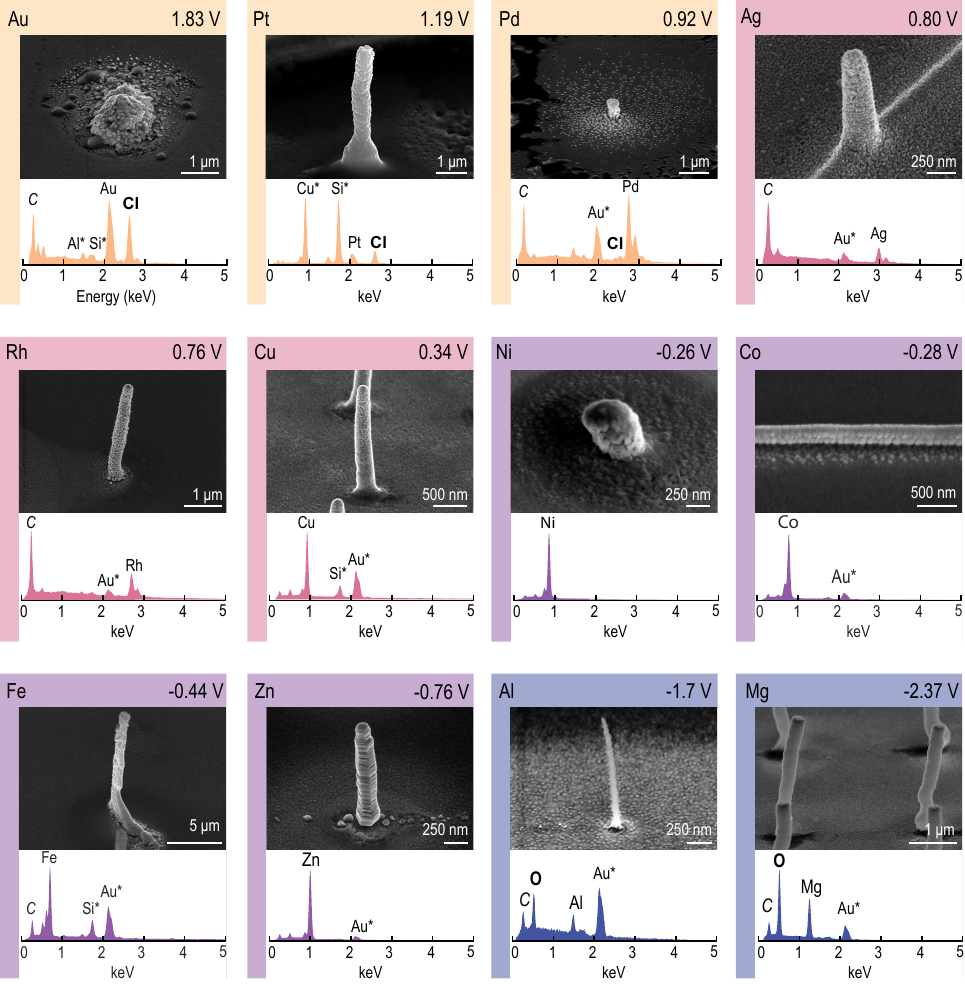}
 \caption{\textbf{Extended range of elements available for deposition in droplet confinement}. Scanning electron micrographs and EDX analyses of elements from a wide range of standard electrode potentials indicated in the top right of each box (relative to the standard hydrogen electrode, SHE). Elements marked with an asterix refer to the used electrode metal (substrate material) and are not part of the deposited structure. Bold indicates non-carbonaceous elements present in significant quantities that are neither the deposited element nor the substrate. Elements shaded on yellows contained traces of counter-ions or complexing elements, namely chloride for the utilized chloride salts. Note that Au could only be ejected by applying a negative potential, presumably due to the negatively charged Au tetrachloride ion. Elements shaded in red allowed for a facile deposition, while elements shaded in purple necessitated a pH between 3 \textendash 5 to enable a deposition of a pure structure. Elements shaded in blue contained a significant amount of oxygen. The prevalence of carbon does not correlate with the electrode potential of the deposited material.}
  \label{Printed_elements}
\end{figure*}

\subsection{Droplet confined deposition of pure metals}
Previous works on droplet-confined electroplating focused on the deposition of Cu, Ag, and Zn \cite{Reiser2019Multi-metalScale, Nydegger2022AdditiveStructures, Porenta2023Micron-scalePrinting}. Here, we use 1~mM solutions of inorganic salts dissolved in water as electrolytes for the deposition of a selection of different elements (Figure~\ref{Printed_elements}). Usually, a chloride or sulfate salt has been selected (as described in the experimental section). Each SE image of an element is accompanied by an EDX analysis of their chemical composition. The underlaid colors group elements by commonality in deposition behavior: For noble metals like Au, Pt, and Pd (shaded in yellow) the co-deposition of chloride is observed. For Ag and Cu (shaded in pink) no deposition of the counter-ions is observed when deposited from 1~mM solutions (both Cl and SO\textsubscript{4} salts have been used). For Rh, a co-deposition of the counterion is difficult to verify as the Cl peak overlaps with the Rhodium peak in EDX. Nevertheless, a defined Rh structure was deposited, although with a high carbon content. Elements with a standard electrode potential between 0 and -0.82~V (Ni, Co, Fe, and Zn, shaded in purple) required the addition of an acid (commonly HCl or HNO\textsubscript{3} to a pH of 3-5) to enable the deposition of metallic structures.\cite{Nydegger2022AdditiveStructures} The influence of the pH is further illustrated in the supplementary information (Figure S1), where pH-dependent EDX spectra are shown.

\begin{figure}
    \centering
    \includegraphics[width= \columnwidth]{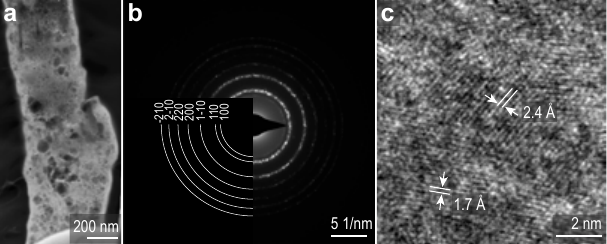}
    \caption{\textbf{Analysis of deposited magnesium. (a)} Dark field TEM reveals a very porous microstructure of deposited magnesium compounds. \textbf{(b)} Selected area electron diffraction reveals a ring-like pattern that fits well with the expected pattern for nanocrystalline cubic MgO (spacegroup F$m\overline{3}$m). \textbf{(c)} The presence of a nanocrystalline material is further supported by high resolution TEM.}
    \label{MgOanalysis}
\end{figure}

It is known that elements with a standard electrode potential below the hydrogen evolution reaction at -0.82~V (Al and Mg, shaded in blue) cannot be electroplated in metallic form from an aqueous solution under normal conditions, as the water will be reduced preferably over the metal-ions in solution. The cations will subsequently precipitate as hydroxides. Indeed, EDX analysis in Fig. \ref{Printed_elements} shows high oxygen contents for Al and Mg pillars, as would be expected from hydroxides or oxides. The pillars appear as homogeneous white structures with no surface patterning and streaking of the image around the pillars' periphery. These streaks are most probably due to charging of the structures under electron beam imaging and indicate poor conductivity. Dark field transmission electron microscopy (TEM) reveals a highly porous structure (Fig. \ref{MgOanalysis}). Electron diffraction indicates the presence of nanocrystalline MgO. The ring patterns match well with the pattern expected for cubic MgO (spacegroup F$m\overline{3}$m). High resolution TEM confirms the presence of a nanocrystalline microstructure (Fig.\ref{MgOanalysis}c).  

\subsection{Towards exotic materials: the curious case of Nickel Phosphorus Oxide \label{NiPO Results}}

\begin{figure}
    \centering
    \includegraphics[width= \columnwidth]{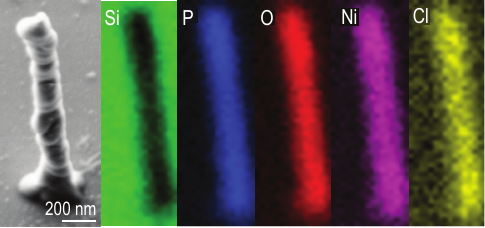}
    \caption{\textbf{Deposition and chemical analysis of Ni-P-O system.} Deposited pillar from a mixture of an aqueous NiCl\textsubscript{2} salt solution and an aqueous hypophosphoric acid solution (H\textsubscript{3}PO\textsubscript{2}) with EDS showing presence of P (blue), O (red), Ni (pink), and Cl (yellow) in the final structure and the Si (green) of the substrate.}
    \label{fig:Ni-P-O-Cl}
\end{figure}

The co-deposition of counter-ions observed for noble metals was also previously observed for Cu \cite{Porenta2023Micron-scalePrinting} and depended on the metal-salt concentration. While co-deposition is undesirable for the deposition of pure metals, it can be leveraged to allow for intentional incorporation of negatively-charged species to create more chemically complex compounds. For example, introducing phosphorus ions by mixing a hypophosphoric acid (H\textsubscript{3}PO\textsubscript{2}) with a nickel salt solution results in co-deposition of phosphorus species and the nickel salt anion (\textit{e.g.}, Cl\textsuperscript{-}) into the printed Ni-based structures (Figure \ref{fig:Ni-P-O-Cl}). However, in previous work by Porenta \& Nydegger \cite{Porenta2023Micron-scalePrinting}, high ion concentrations ($\geq$ 10~mM) inhibit the formation of pillar and 3D structures and result purely in broad, planar deposits containing both the metal ion and its counter ions. In contrast, for this Ni-P-O system (Figure~\ref{fig:Ni-P-O-Cl}), well-resolved pillar structures are deposited even with similarly high ion concentrations, and the chloride or phosphor-containing anions are incorporated directly into the structures. 

As noted above for Ni deposition, the pH must be low for direct metal deposition. Although these solutions ranged from a pH of 2.55 to 2.75, depending on the Ni:P ratio, all printed structures contain strong oxygen signatures. This is in contrast to the pure NiCl\textsubscript{2} solution, which produces pure Ni at similar acidity. Due to the more complex deposition process with the added P and oxygen source from the hypophosphoric acid, this low pH may not be enough to inhibit hydroxide and oxide formation.

\subsection{Combining the deposition of different material classes} 

\begin{figure*}[t]
    \centering
    \includegraphics[scale = 1]{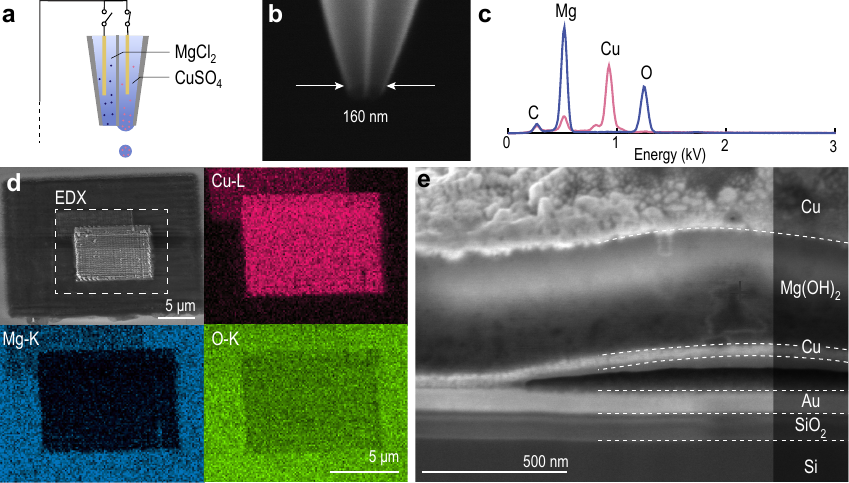}
    \caption{\textbf{Co-deposition of a metal (Cu) and a dielectric material (Mg(OH)\textsubscript{2}). (a-b)} A two channel nozzle as utilized to demonstrate the multi-material-classes deposition. One channel is filled with 1~mM CuSO\textsubscript{4} and the other with 1~mM MgCl\textsubscript{2}. Best deposition results were obtained for orifice sizes around 160~nm.\textbf{(c)} EDX analysis of a Cu and Mg(OH)\textsubscript{2} pad indicates that the materials can be deposited separately from the same nozzle with almost no intermixing (a very small Mg signal is observed in the Cu pads). \textbf{(d)} Low magnification image of a 5 by 5 \textmu m Cu pad on top of 10 by 10\textmu m magnesium hydroxide pad. The inset is shown in the EDX maps. The spatial distribution of Cu, Mg, and O indicates that Cu is only present in the square where it was deposited (the square at the top left in the Cu image originates from a shift in the position of the substrate during printing) and that Mg is mostly found in the larger pad surrounding the Cu. \textbf{(e)} Cross-section of a Cu-Mg(OH)\textsubscript{2}-Cu multi-layered structure. The first Cu layer delaminated from the Au substrate. Both Cu layers appear much brighter than the porous Mg(OH)\textsubscript{2}. Further, charging of the Mg(OH)\textsubscript{2} layer can be observed as a brighter section within the layer.}
    \label{CuonMgO}
\end{figure*}

Based on the promising initial results of the deposition of both metallic and oxidic monolithic structures, we extend the deposition towards metal-ceramic composite structures. Hereto, we utilize two-channel nozzles as described previously\cite{Reiser2019Multi-metalScale}, where one channel is filled with 1~mM CuSO\textsubscript{4} and the other with 1~mM MgCl\textsubscript{2} (Fig. \ref{CuonMgO}a-b). 

An EDX analysis of the materials printed separately (Spectra shown in Fig. \ref{CuonMgO}c, pads not shown) confirms that the two channels allow the deposition of the two materials separately with little intermixing (only a very small Mg peak is observed in the Cu section and vice versa). Fig. \ref{CuonMgO}d shows an SEM image depicting a Cu pad (brighter square in the middle) on top of a larger black square, which demonstrates that these structures can be printed on top of each other, hence effectively directly depositing multi-material-classes structures (\textit{i.e.} metal and ceramics in the same structure). The magnesium oxide square is larger than the Cu pad to ensure there is no connection between the Cu pad and the conductive Au substrate. EDX maps, taken of the smaller area depicted in Fig. \ref{CuonMgO}d confirm that Cu is only deposited in the square pattern (the copper on top left is due to a shift of the nozzle position during printing). The larger black pad is confirmed to consist mainly of Mg and O. A cross-section of a Cu-Mg(OH)\textsubscript{2}-Cu pad (Fig. \ref{CuonMgO}e) also demonstrates that the layers can be stacked (although the bottom Cu layer delaminated from the Au substrate). It can also be seen that the Mg(OH)\textsubscript{2} layer is porous, and the top Cu layer exhibits a high roughness.

\section{Discussion}
\subsection{The governing electrochemical principles in droplet-confined electroplating}
The elements and materials for which we demonstrated deposition are by no means a comprehensive list. We have therefore derived a framework to serve as a starting point for future work to deposit further materials. The principles that govern the plating in droplet-confinement are primarily due to the unique electrochemical setting of the technique, namely the unidirectional ejection of cation-loaded droplets and the reduction within transient solvent droplets (\textit{i.e.}, all solvent evaporates intermittently or after the deposition), which prevents ion exchange with the larger electrolyte reservoir in the nozzle\cite{Menetrey2024NanodropletPrinting, Nydegger2024Droplet-ConfinedNanowires}. The present work focuses on utilizing water as a solvent and electrolyte, although the found principles may apply analogously to other solvents. We use water as a model system here due to the extensive body of literature available for classical electroplating in water\cite{2010ModernElectroplating, Schwartz2010DepositionOverview}. However, the comparison to classical electroplating is complicated by the fact that the conditions (pH, metal ion concentration, temperature) within the droplets in EHD-RP are highly dynamic and not known. Therefore, we have to resort to more general statements that we base both on observations and theoretical arguments. We discuss first the requirements for metal plating in the droplet, and based on these requirements, derive the requirements for the ion generation mechanisms. 

\begin{figure}[!htbp]
    \centering
    \includegraphics[width = \columnwidth]{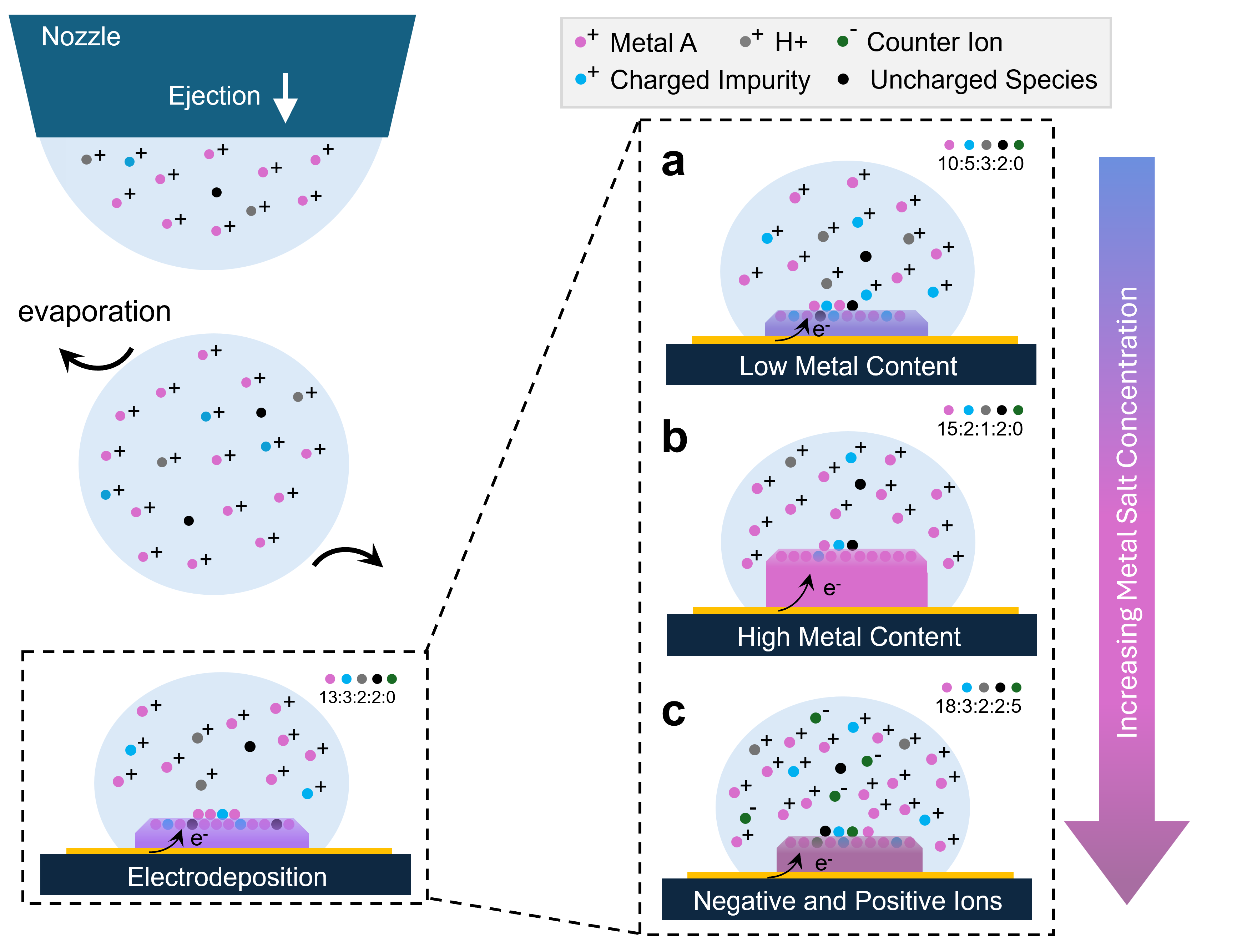}
    \caption{\textbf{The importance of the charge carriers on the purity of the deposited structures.} Schematic of ejected nanodroplets with a focus on the charge carriers, including the depiction of dissolved metal A (pink), charged impurities (blue), H\textsuperscript{+} ions (grey), negatively charged counter ions (green), and uncharged species (black). Note that the net charge of the droplet is conserved from ejection until reduction on the substrate or until the droplet disintegrates \cite{Menetrey2024OnPrinting}. The same net charge of the droplet can be achieved with various combinations of charge carriers. \textbf{(a)} a low metal content and high content of H\textsuperscript{+} or charged impurities lead to a slow deposition of metal structures with large impurity incorporation. \textbf{(b)} high metal content and a minimal presence of H\textsuperscript{+} and charged impurities leads to the highest deposition rate. \textbf{(c)} At very high metal-ion concentration, negatively charged counter-ions are co-ejected and subsequently co-deposited. Count of specific ions are to the right of each nanodroplet, and for all droplets the net charge is equal.}
    \label{Fig: Discussion}
\end{figure}

\subsubsection{Reduction in the droplet on the substrate}

\textit{Reactivity of the metal}\\
An obvious requirement for the deposition of a metallic structure is that the metal ion must preferably be reduced over the solvent on the substrate, which serves as the cathode. In the case of water as a solvent, this information is already compiled as the standard electrode potential for standard conditions (25\degree C, 1~M concentration for solutions, 1~atm pressure for gases). To deposit a metal from an aqueous solution, the reduction of the metal must take place preferentially over the cathodic partial reaction of water splitting (\ce{2 H2O -> 2 H+ + 2OH-} E\degree : -0.8277~V versus standard hydrogen electrode, SHE\cite{Haynes2014}). A metal with a more negative electrode potential would lead to hydrogen formation and the precipitation of the metal as a hydroxide (M(OH)\textsubscript{x}). This requirement is illustrated by the fact that metallic Zn (E\degree : -0.7618~V\cite{Haynes2014}) was successfully deposited from aqueous acidic solutions, while magnesium and aluminum yielded oxygen-rich structures. 

\textit{Counter-ions and other ligands in the solvent droplet}\\
Since the droplet is not connected to a liquid reservoir that could allow for ion exchange, all non-volatile species present in the droplet are ultimately deposited when the solvent evaporates. Therefore, we refer to all unwanted and non-volatile species as contaminants. A notable source of contaminants are the counter-ions and other ligands of the metal cation. If instead of cation-only loaded droplets also counter-ions are ejected, partially reduced metal salts are deposited. This was previously observed for a 10~mM CuSO\textsubscript{4} solution \cite{Porenta2023Micron-scalePrinting} and in this work also for noble metals at 1~mM concentration (~\ref{Printed_elements}). 

We base this finding on the presence of complexes of the metal salts that are not fully dissociated. In fact, speciation diagrams imply that many metals ions form complexes with counter-ions at concentrations at or above 1~mM \cite{Powell2007ChemicalReport, Powell2013ChemicalReport}. Such complexes can be detected with electrospray mass spectrometry (shown in previous work for Zn and NO\textsubscript{3} complexes\cite{Nydegger2022AdditiveStructures}). Especially noble metal chlorides are known to form highly stable, partially covalent bonds \cite{Wojnicki2012KineticAgent, Mahlamvana2014PhotocatalyticAgent, VanMiddlesworth1999TheBar}. Here, very low concentrations ($<<$ 1~mM) would have to be used to form fully solvated metal ions. In fact, chloride was always detected (with EDX) in Au, Pd and Pt structures deposited from the respective chloride salts at 1~mM concentration (Fig. \ref{Printed_elements}). Therefore, a second requirement for the deposition of pure metal structures can be determined: only fully solvated cations should be ejected. 

Complexes with a net-charge of zero (such as CuSO\textsubscript{4}) should not be attracted towards the apex of the nozzle by the electric field, but are homogeneously distributed in the electrolyte, and are hence also co-ejected. Partially charged species, such as CuCl\textsuperscript{+} should also be attracted to the nozzle and are frequently co-ejected. Notably, the requirement that all cations should be fully solvated also excludes ligands that are typically used to improve solubility of metal cations, such as ethylenediaminetetraacetic acid (EDTA). This chelating ligands usually form highly stable complexes leading to co-ejection of complexing agents~\cite{Boija2014DeterminationSpectrometry}. 

\textit{Towards contamination-free metal structures}\\
Also the third requirement for the deposition of pure metals emerges from the fact that all non-volatile species present in the droplet are deposited. These non-volatile species include unwanted positively charged ions (such as Na\textsuperscript{+} and Ca\textsuperscript{2+}), but unfortunately also any other non-volatile species present in the electrolyte, which are not repulsed from the orifice by an applied electric field (like negatively charged counter-ions) and are therefore co-ejected. A high purity of the deposited metal structure can be achieved by ensuring a high ratio between metal ions and other non-volatile species in the droplet. In previous studies, metal-ion concentrations of 0.1~mM and below resulted in carbon, sodium and calcium-rich deposits\cite{Porenta2023Micron-scalePrinting}. The origin of the contamination has so far not been traced. Likely, the purity of the deposited metal structures is related to the ratio of metal-ions to contaminants in the residing droplet on the substrate (Fig. \ref{Fig: Discussion}). The importance of this metal-to-contaminant ratio therefore naturally implies two possible directions of work towards purer metal structures: either the amount of carbon and other impurities has to be reduced or the amount of metal cations present in the droplet has to be increased.

First, the origin of carbon is currently not known, but should be traced and avoided. So far, neither the use of an O\textsubscript{2} plasma cleaner to clean all glassware before use nor ultrapure solvent brought significant and reliable improvements in the observed level of carbon in the structures. Here, an interesting approach could be to utilize electrospray mass spectroscopy to further investigate the nature of the carbon impurities, with the goal of determining its origin. 

Second, the concentration of metal cations in the droplets should be maximized. As the ejection process is charge driven\cite{Menetrey2024NanodropletPrinting}, monovalent cations allow to increase the number of metal-atoms while maintaining the same overall charge and should therefore be preferred. Yet, many metals form bivalent species in water\cite{Powell2007ChemicalReport, Powell2013ChemicalReport}. Another contribution to the overall charge comes from H\textsuperscript{+} that is present in the solution (which explains the lower-than-expected Faraday efficiency observed in previous work for aqueous solvents\cite{Nydegger2024Droplet-ConfinedNanowires}). The role of H\textsuperscript{+} is ambiguous, however: On the one hand, it lowers the amount of metal-cations per droplet due to its charge. This probably reduces the efficiency of the deposition (more droplets are necessary for the same volume of metal deposited), which, in combination with the assumption of a constant background level of contamination, leads to less pure metals at lower pH. For Ni and Fe, on the other hand, it helped improve the deposition, but it can lead to porosity in the deposited structure. A hypothesis is that the addition of H\textsuperscript{+} suppresses the precipitation of Ni and Fe hydroxides. The formation of hydroxides is a known phenomena in electroplating, sometimes observed at locations of high current density (\textit{i.e.}, edges)\cite{Aerts2010AnodizingDensities, 2010ModernElectroplating}. Here, after a local depletion in metal cations, the electrolysis of water leads to formation of hydrogen gas and causes a local rise in the pH. This rise in pH then leads to the precipitation of metal hydroxides. For droplet confined electroplating, the high current densities (0.6–1.8~A cm\textsuperscript{-2} \cite{Menetrey2022TargetedPrinting, Nydegger2024Droplet-ConfinedNanowires}) and a local electron surplus (provided by a capacitive effect of the setup) might lead to the onset of water electrolysis. Hence, H\textsuperscript{+} are  enabling the deposition of metallic structures for elements with an electrode potential between the cathodic partial reaction of water splitting and the reduction of hydrogen ( -0.8277~V $<$ E\degree $<$ 0~V) \cite{Nydegger2022AdditiveStructures}.

\subsubsection{Ion generation in the nozzle: Choosing metal salts and electrode materials}
The observed contamination in the deposited structures implies a minimal cation concentration and therefore a minimal metal salt solubility of 0.1~mM in water, albeit the solubility is ideally above 1~mM for purer structures. Notably, however, these metal cations can be of different elements. For example, a mixture of CuSO\textsubscript{4} and ZnSO\textsubscript{4} or AgSO\textsubscript{4} leads to the deposition of an alloy with equal stoichiometry to the utilized solution. Mixed salt solutions also allow for the fabrication of equiatomic Ag-Cu-Zn alloys. It is important for the deposition of alloys that compatible precursor salts are chosen. For example, a copper chloride precursor might be soluble, but the chloride anion would precipitate with any silver-ions present due to the low solubility product of AgCl. Ideally, non-coordinating anions such as perchlorates are chosen.

Further, an immersed electrode is necessary to establish the electric field for the ejection and to regenerate the charges extracted at the tip. Without the formation of positive charges at the anode, the extraction of positive charges at the tip would lead to a net charge of the nozzle. This would then lead to an electric field counteracting the field used for droplet ejection and stop the ejection when the two contributions cancel out each other. Therefore, the anode should oxidize the solvent without passivating itself. To avoid the formation of a secondary metal ions by anodic dissolution a non-dissolving, non-passivating electrode should be used. The standard for droplet-confined electroplating is currently the use of an Au electrode when salt solutions are printed. The Au electrode allows for the oxygen evolution reaction in the nozzle, to ensure charge neutrality by the formation of H\textsuperscript{+} ions (E\degree: 1.229~V\cite{Haynes2014}).\\ 

To summarize, the following set of requirements can be regarded as necessary for the deposition of pure metals in the confinement of transient water droplets:\\

\textit{1: The concentration-adjusted standard electrode potential of the to-be-deposited metal is higher than -0.829~V.}\\

\textit{2: The metal ion forms a fully solvated species in the droplet.}\\

\textit{3: Concentration of metal cations in H\textsubscript{2}O $\geq$ 0.1~mM, given by the background level of contaminants and acidity (concentration of H\textsuperscript{+}) in water.} \\

\textit{4: Mixed metal salts require sufficient solubility ($\geq$~0.1~mM) for all combinations of cations and anions.} \\

\textit{5: The electrode is made of a metal that has a standard electrode potential $>$ 1.229~V}\\

\subsection{Direct deposition of insulators}
The first open question for structures deposited from Mg and Al solutions concerns the actual nature of the deposited material—whether it is a metal, a hydroxide, or an oxide. We currently hypothesize that the difference between literature (predicting Mg(OH)\textsubscript{2})\cite{Shinagawa2021} and electron diffraction (indicating MgO) is explained by a dehydroxylation of Mg(OH)\textsubscript{2}\cite{Gomez-Villalba2016TEM-HRTEMOxide} when exposed to an electron beam (or high-intensity lasers) during analysis. Therefore, we assume that metal hydroxides are deposited, however we note that further careful analysis of deposited structures is necessary.  

The second open question is how confined out-of-plane deposition of tall, insulating metal hydroxides is possible, since the reduction of water on the substrate requires the presence of electrons. Here, one could entertain two possible reaction mechanisms: first, a water bridge connects the growth-front with the conductive substrate and enables the formation of hydroxide on the substrate, which later precipitates with incoming Mg cations. Second, and more probably, is that the charges can pass through the magnesium hydroxide itself. The very high applied electric fields, the porous nature of the hydroxide and the arguably high level of contamination implies that the material does not act as a perfect insulator. Electron transfer through the deposited material would also explain the presence of metallic copper on-top of deposited Mg(OH)\textsubscript{2} as shown in Fig. \ref{CuonMgO}. Interestingly, the deposited material appears as non-conductive under the electron beam (indicated by the horizontal streaks in many SEM images) in Fig. \ref{CuonMgO}. This could be related to different strengths of electric fields during deposition and analysis or that the low current during deposition could tunnel through the insulator (the deposition utilized a current of 1~nA while SEM was carried out with a current of 0.1~nA, but the e-beam is much more focused than the deposition, hence the deposition had a lower current density).

Future experiments should determine the conductivity or the breakdown electric fields of the deposited Mg(OH)\textsubscript{2}, as this information is crucial to determining the reaction path. Further, UV-photoluminescence\cite{Kumari2009SynthesisMgO} could give information about the insulating behavior of the deposited material (although care has to be taken to avoid an annealing of the material with the laser beam), as this technique is, in principle, sensitive to electronic states of defects within the band gap. Lastly, an experiment determining the maximal height of printed Mg(OH)\textsubscript{2} structures would be important. Future developments could use a dehydroxylation under irradiation to optimize the deposition towards dense oxides. Here, one could combine the deposition of Mg(OH)\textsubscript{2} with \textit{in situ} annealing using either an electron beam or a focused laser-beam, as has been used in previous studies to heat up the substrate locally \cite{Nydegger2024DirectSize,Menetrey2023FlyingPrinting}.

\subsection{Case Study for NiPO}
Based on the findings of metallic and insulator deposition, the Ni-P-O system is presented as a case study to explore the deposition of functional materials with increased chemical complexity. Specifically, doped transition metal oxides, such as phosphorus-doped NiO\textsubscript{x}, represent a class of tunable wide-bandgap semiconductors that have promising applications in photonics \cite{WBS_photonics,NiOP_photonics} and integrated electronics\cite{WBS_electronics,NiPO_electronics}. The results from Section~\ref{NiPO Results} already highlight the versatility of nanodroplet-confined electrodeposition as Ni-P-O material systems are not typically produced through electrochemical methods. Furthermore, electrodeposition of Ni-P alloys often involves a two-step process at elevated bath temperatures as discussed in the Supplementary Information. These factors point to the highly kinetic nature of deposition in these confined nanodroplets and a new opportunity for the design of chemically complex materials.

\begin{figure}
    \centering
    \includegraphics[width= \columnwidth]{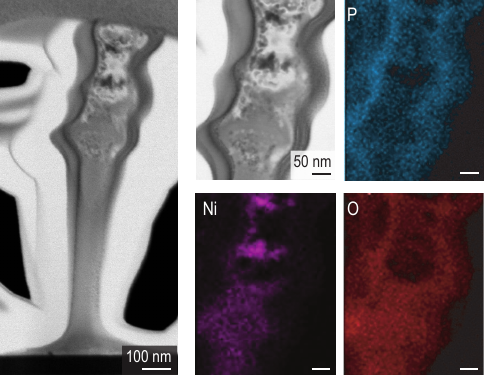}
    \caption{\textbf{Dark field transmission SEM image of a deposited pillar}. The pillar is deposited from an aqueous hypophosphoric acid solution (H\textsubscript{3}PO\textsubscript{2}) with Ni\textsuperscript{2+} ions produced through anodic dissolution of a Ni wire. Zoomed in section of pillar top shows EDS mapping of P (blue), Ni (pink), and O (red).}
    \label{fig:Ni-P-O transmission}
\end{figure}

However, challenges remain with incorporating a negatively charged dopant ion, such as P, without incorporating the metal salt's cation (Figure \ref{fig:Ni-P-O-Cl}). To circumvent this issue, Ni\textsuperscript{2+} ions are introduced through electrocorrosion of a pure Ni wire in an aqueous hypophosphoric acid solution. (Details on selection of hypophosphoric acid over phosphoric acid are outlined in the Supplementary Information). This produces Cl-free materials that contain only Ni, P and O as shown in Figure \ref{fig:Ni-P-O transmission}.

Both metal salt and corrosive wire ion sources produce pillars with a strong oxygen signal and a clear presence of incorporated P. These structures also exhibit globular surface features that point to a greater fluctuation in droplet-to-droplet deposition compared to the pure chemistry of Ni salts (Figure \ref{Printed_elements}). Transmission SEM images of a cross section of a Ni-P-O pillar highlight the strong presence of oxygen throughout the structure. It reveals limited crystallite formation near the base, although some larger Ni-rich grains have formed near the top (Figure \ref{fig:Ni-P-O transmission}). This points to the deposition of amorphous phosphorus-containing nickel hydroxides. Such hydroxides are observed in Ni salt-based deposition as discussed previously. While the pH of the solution in these depositions are below the observed threshold of a pH between 3 and 5, other kinetic effects due to the presence P or additional reactions at the surface may shift the propensity for hydroxide formation. These deposited Ni-P-O structures also show some Ni-rich crystallites which have formed within the amorphous matrix. While the observed Ni-rich crystallites may form during deposition, beam-induced crystallization either during imaging or ion milling cannot be discounted. 

\begin{figure}
    \centering
    \includegraphics[width= \columnwidth]{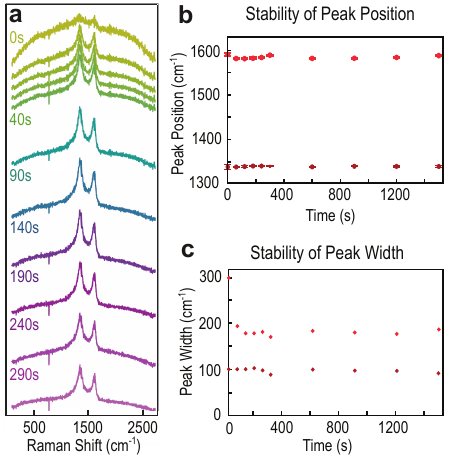}
    \caption{\textbf{Stability of NiPO under laser illumination. (a)} Raman spectra offset to highlight different timesteps in the exposure with 0-40s in 10s intervals and 40-340s in 50s intervals. Dark yellow being the initial timestep and the bright purple being the last timestep. \textbf{(b)} Peak position of both visible peaks over full 25 minutes of exposure and \textbf{(c)} the width of those peaks over the exposure.}
    \label{fig:Ni-P-O Raman}
\end{figure}

Furthermore, the time-resolved Raman spectra in Figure \ref{fig:Ni-P-O Raman} contain a broad initial peak with two sharper peaks at 1339 cm$^{-1}$ and 1585 cm$^{-1}$. The broad peak, which disappears within the first 10 seconds of data collection, could be attributed to an amorphous phase such as an amorphous nickel hydroxide Ni(OH)\textsubscript{2} which is then annealed under the laser.  The Raman spectra after 10 seconds appears similar to other spectra observed in carbon-containing Ni\textsubscript{3} PO\textsubscript{4})\textsubscript{2}\cite{Ni3(Po4)2}. These peaks remain stable over 25 minutes of exposure (Figure \ref{fig:Ni-P-O Raman}). Raman peaks between 1220-1360 cm$^{-1}$ are associated with P-O bonds \cite{PO_Raman} while the second peak at 1585 cm$^{-1}$ is aligned with a nanocrystalline NiO\textsubscript{x} containing high defect densities\cite{NiO_Raman}. This peak aligns well with the NiO\textsubscript{x} 2M peak, indicating magnetic ordering. However, the effects of P on peak position and shape are not well understood and likely contribute to slight deviation in peak position. Coupled with the EDS and microscopy images, these results further point to an initial amorphous Ni(OH)\textsubscript{2} structure which is likely annealed under the Raman laser and reduced into a stable Ni-P-O system. 

This case study highlights the opportunity for design of unusual and chemically complex materials at the nanoscale through confined nanodroplet deposition. By leveraging the presence of counter-ions or negatively charged species, this technique can introduce doping or advanced alloying to materials from ceramics to metals and semiconductors. This Ni-P-O system also highlights the opportunity to mix solution-based ions with ions introduced through electrocorrosion. Such methods are critical when needed to select a specific negatively charged ion species and could be extended to control metal ion incorporation through electrocorrosion.

\section{Conclusion and Outlook}
In summary, we have demonstrated that EHD-RP can be used for the additive manufacturing of a much wider range of metals with submicron resolution than previously demonstrated. Specifically, we investigated different approaches to provide a sufficient cation concentration in the electrolyte. Here, the use of metal salt solutions and diluted acids provided access to Co, Ni, Au, Pd, and Rh structures in addition to the previously reported Zn, Cu, and Ag. 
 
To guide the future search for printable metals, we have attempted to derive the fundamental principles that govern the deposition in transient solvent droplets. A set of 8 rules has been derived, based on both fundamental arguments and experimental results.

However, before reliable deposition of functional structures is attempted, both the purity of the deposits and the reliability of the process must be increased. A major challenge appears to be the optimization of the ratio between ligand-free metal ions and other non-volatile species. Both approaches for ion-sources in the current setting have limitations in achieving a high metal-to-contaminant ratio: metal salts form cation-anion complexes at high concentration, while sacrificial anodes can passivate at the high applied voltages that are necessary to eject the droplets for many elements. Therefore, future work should focus not only on tracking the source of contamination but also on investigating alternative approaches that avoid the passivation of sacrificial anodes.  

\section{Experimental}
\subsection{Materials}
Nozzles for deposition were fabricated on a P-2000 micropipette puller system (Sutter Instruments) from filamented Quartz capillaries (Sutter Instruments, Item QF100-70-15). Nozzle diameters were determined on a Quanta 200F (Thermo Fisher Scientific), equipped with a Schottky type field emission gun (FEG) in low vacuum mode (30~Pa) to avoid charging. Nozzles with diameters of 60–90~nm for single channel and 160~nm for dual-channel were used. Au wires (0.25~mm Metaux Precieux SA, 99.999\%) were used as ejector-electrodes.

For the preparation of aqueous metal salt solutions, high-purity water (LC/MS-Grade, Fisher Chemical)) and the following metal salts were used: \ce{CuSO4 . 5 H2O} (Sigma Aldrich, 99.999\% metal basis), \ce{CuCl2 . 2 H2O} (Sigma Aldrich, 99.999\% metal basis), \ce{Ag2SO4} (Alfa Aesar, 99.999\% metal basis), \ce{ZnCl2} (Alfa Aesar, 99.995\% metal basis), \ce{CoCl2 . 6 H2O} (Sigma Aldrich), \ce{HAuCl4 . 3H2O} (Sigma Aldrich,  $\geq $ 49.0\% metal basis), \ce{PdCl2} (Fluorochem Ltd), \ce{RhCl3 . 4H2O} (99.99\% metals basis, Alfa Aesar), \ce{FeSO4 . 7 H2O} (Fluka Analytical), \ce{NiCl2} (Sigma Aldrich, 99.999\% metal basis). All chemicals were used as received. The nozzles were filled by using gas-tight glass syringes with custom made PEEK tips. The nozzles were cleaned on the inside by rinsing with the solvent. Substrates were 0.4~cm × 2.0~cm pieces of Au coated (80~nm) Si wafers. The substrates were cleaned in technical acetone, analytical isopropanol, and subsequently blow-dried with compressed air before use. Cu wires (Alfa Aesar, 0.25~mm diameter, 99.999\% metal basis) were etched in concentrated nitric acid (65\% HNO3, Sigma Aldrich) for 10~s, then dipped in water and stored in analytical ethanol until use (maximal 10 minutes), while Au wires (Metaux Precieux SA, 99.999\%) were immersed for 30~s, dipped in water and stored in air.

For the salt-based ion source in the Ni-P-O experiments, an aqueous solutions with a 10~mM total ion concentration was mixed using a 2:3 Ni:P ratio by using \ce{NiCl2} (Sigma Aldrich, 99.999\% metal basis), hypophosphoric acid (H\textsubscript{3}PO\textsubscript{2} ; Sigma Aldrich) and high-purity water (LC/MS-Grade, Fisher Chemical). 

For wire corrosion ion source in the Ni-P-O experiments, ions were produced from a Ni wire (Alfa Aesar, Premion, 99.999\% metal basis) in a 10~mM hypophosphoric acid (H\textsubscript{3}PO\textsubscript{2} ; Sigma Aldrich) aqueous solution made using high-purity water (LC/MS-Grade, Fisher Chemical)

\subsection{Deposition}
Deposition was performed at room temperature in an argon atmosphere ($<$100~ppm O\textsubscript{2}). Typical potentials applied to the anode during printing were 80\textendash 160~V. For printing with a certain current the potential was adjusted to match this ejection current by averaging the current over 40 data points (spaced 0.2~s) and adjusted in 0.2~V increments if the average deviated more than 5\%. Nozzle substrate distance was controlled to be 5\textendash 15~\textmu m by analysing the light microscope images and adjusting the position of the z-axis piezo stage accordingly.
 
\subsection{Analysis} 
High Resolution Scanning electron microscopy (SEM) was performed with a Magellan 400 SEM (Thermo Fisher Scientific, former FEI) equipped with an Octane Super EDXsystem (EDAX, software: TEAM). Tilt angles were 55\textdegree. HR-SEM images were taken in immersion mode with an acceleration voltage of 5~kV. A dual-beam Helios 5UX (Thermo Fisher Scientific) with a focused Ga\textsuperscript{+} liquid metal ion source was used for focused ion beam (FIB) milling of  cross-sections. Prior to FIB-milling the pillar was coated by a protective carbon layer, which is visible in the SEM image as a black background. Transmission mode SEM imaging and EDS was also performed in the dual-beam Helios 5UX.

\section{Author Contributions}
M.N. devised the concept to deposit both metallic and oxidic materials from salt solutions, while R.A.G. devised the concept to deposit Ni-P-O. A.B. and R.A.G. deposited and analyzed Ni-P-O. M.N. and R.A.G. visualised the data and wrote the original paper draft. M.N., R.G., and R.S.  discussed the results. R.S. supervised the project and all authors reviewed the manuscript.

\section{Conflicts of interest}
The authors declare no conflicts of interest.

\section{Acknowledgement}
This work was funded by grant no. SNF 200021-188491. The authors thank Tingyi Wang, Raffael Adamek, Pauline Fichter, Xuanlin Peng, Tristan Sachsenweger, and Fabian Haake for additional deposition experiments. Electron-microscopy analysis was performed at ScopeM, the microscopy platform of ETH Zürich.

\bibliography{Manual_references} 
\bibliographystyle{rsc} 

\newpage
\newcommand{\beginsupplement}{%
        \setcounter{table}{0}
        \renewcommand{\thetable}{S\arabic{table}}%
        \setcounter{figure}{0}
        \renewcommand{\thefigure}{S\arabic{figure}}%
     }

\beginsupplement
\onecolumn
\noindent\Large{\textbf{Supplementary Information}}\\
\\
\\
\noindent\LARGE{\textbf{Nanodroplet-Confined Electroplating Enables Submicron Printing of Metals and Oxide Ceramics}} \\
\\
\noindent\large
\large Mirco Nydegger\Mark{1,2}$^\dag$,
        Rebecca A. Gallivan\Mark{1,3,}$^\dag$$^{*}$,
        Arthur Barras\Mark{1}
        and Ralph Spolenak\Mark{1,}$^{*}$
       \\[1em]

\normalsize
 
\Mark{1}\small{Laboratory for Nanometallurgy, Department of Materials, ETH Zürich, Vladimir-Prelog-Weg 5, 8093 Zürich , Switzerland}\\ 

\Mark{2}\small{Present address: Department of Aeronautics and Astronautics, MIT, 77 Massachusetts Avenue, Cambridge, MA 02139, USA}\\ 

\Mark{3}\small{Present address: Thayer School of Engineering, Dartmouth College, 15 Thayer Drive, Hanover, NH 03755, USA}\\ 
 
$^\dag$These authors contributed equally\\
  
$^{*}$To whom correspondence should be addressed; e-mail: rebecca.gallivan@dartmouth.edu; ralph.spolenak@mat.ethz.ch\\[0.5em]
 
\normalsize

\vspace{0.5cm}
\setcounter{page}{1}
\setcounter{section}{0}

\section{Setup}

Fundamentally, the setup consists of piezo stages that move the substrate in X, Y, and Z direction (QNPXY-500, QNP50Z-250, Ensemble QL controller, Aerotech). Additionally, stage translations in X and Y direction larger than 500~\textmu m were enabled by additional long-range stages (M112-1VG, PI for Y direction, manual micrometre screw, Mitutoyo for X direction). Piezo stages and power source were controlled through a custom Matlab script. The nozzle is mounted on a motorized nozzle holder (Z825B, controlled with Kinesis, both Thorlabs). The deposition is observed trough an optical microscope composed of a ×50 objective lens (LMPLFLN, Olympus) and a CMOS camera (DCC1545M, Thorlabs). The lens is mounted at an inclination of 60 \degree to the substrate normal and the substrate is illuminated from the behind using a green LED (LEDMT1E, Thorlabs). The green light ($\lambda = 530~nm$) and a $NA = 0.5$ yields a resolution limit for the optical system of 530~nm. A power source (B2902a or B2962a, Keysight) with triaxial cable connectors was used for polarizing the anodes. Up to 4 channels can individually be supplied with high potential through a custom designed PCB board, which utilizes reed-relays to open and close electrical connections (PICKERING, 131-1-A-5/1D). The metal wires were connected using a mechanical clamp. The complete printing setup is mounted inside a custom built gas tight box to provide an oxygen-free atmosphere (controlled with an oxygen sensor, Module ISM-3, Dansensor) and is placed on a damped SmartTable (Newport) to provide a vibration-free environment.

\section{pH dependent EDX}

\begin{figure}
    \centering
    \includegraphics{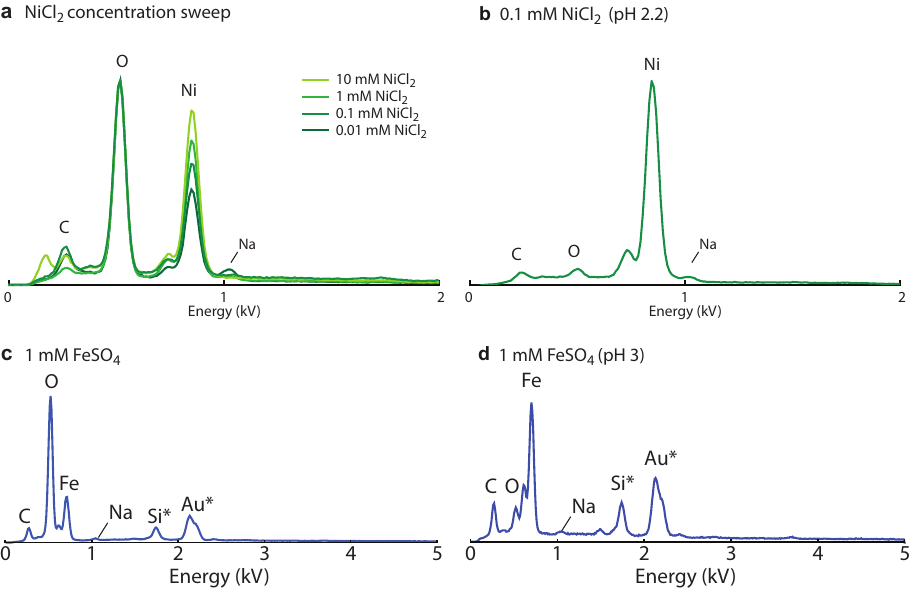}
    \caption{\textbf{Influence of the concentration and the pH on deposited material. (a)} A higher Ni concentration leads to a higher Ni to O ratio. \textbf{(b)} A pH of 2.2 suppresses the Oxygen signal almost completely. \textbf{(c-d)} For Fe, a similar reduction of the O peak can be observed with a low pH in the electrolyte.}
    \label{NiFe_pH}
\end{figure}

The O-to-Ni ratio depends on the NiCl\textsubscript{2} concentration in the electrolyte (Appendix Fig. \ref{NiFe_pH}a). A higher NiCl\textsubscript{2} concentration led to a lower O-to-Ni ratio (however, at 10~mM a Cl signal was found in the EDX analysis). The lowest O content was found when the pH was lowered (using sulfuric acid) to 2.2 (Appendix Fig. \ref{NiFe_pH}b). With such a low pH, very slow deposition of structures was observed (20 \textendash 30 s for the structure depicted in Fig. \ref{Printed_elements}b), but EDX analysis revealed a much lower oxygen content and the surface shows a different morphology. Similar results to Ni were obtained for Fe, which was deposited from 1~mM FeSO\textsubscript{4} solutions. Here, for a non-adjusted pH, structures with high oxygen content were deposited, while the addition of sulfuric acid (pH 3) enabled the deposition of Fe with a lower oxygen content (Appendix Fig. \ref{NiFe_pH}c-d).

\section{Choice of Acid}
Ni-P compounds have previously been electrodeposited from a water-based electrolyte solution using either phosphorous acid (H\textsubscript{3}PO\textsubscript{4}) or hypophosphoric acid (H\textsubscript{3}PO\textsubscript{2})\cite{NiP-Table}. The deposition of Ni-P can occur via either a direct or indirect mechanism of reducing P from a phosphorous-containing acid\cite{NiP-directIndirect, NiP-directIndirect}. Table \ref{Table NiP Half Reaction} highlights both routes in the half reactions that precede Ni-P electrodeposition.

\begin{table}[ht]
\centering
\begin{tabular}
{@{} p{0.75\linewidth} c @{}}
\multicolumn{2}{c}{\textbf{Electrochemical reactions of components in bath for Ni-P plating}} \\[6pt]
$2\mathrm{H}^{+} + 2e^{-} \rightleftharpoons \mathrm{H}_2$ & (1)\\
$2\mathrm{SO}_4^{2-} + 4\mathrm{H}^{+} + 2e^{-} \rightleftharpoons \mathrm{S}_2\mathrm{O}_6^{2-} + 2\mathrm{H}_2\mathrm{O}$ & (2)\\
$\mathrm{Ni}^{2+} + 2e^{-} \rightleftharpoons \mathrm{Ni}$ & (3)\\
$\mathrm{H}_3\mathrm{PO}_4 + 2\mathrm{H}^{+} + 2e^{-} \rightleftharpoons \mathrm{H}_3\mathrm{PO}_3 + \mathrm{H}_2\mathrm{O}$ & (4)\\
$\mathrm{H}_3\mathrm{PO}_3 + 3\mathrm{H}^{+} + 3e^{-} \rightleftharpoons \mathrm{P} + 3\mathrm{H}_2\mathrm{O}$ & (5)\\
$\mathrm{H}_3\mathrm{PO}_3 + 2\mathrm{H}^{+} + 2e^{-} \rightleftharpoons \mathrm{H}_3\mathrm{PO}_2 + \mathrm{H}_2\mathrm{O}$ & (6)\\
$\mathrm{H}_3\mathrm{PO}_2 + \mathrm{H}^{+} + e^{-} \rightleftharpoons \mathrm{P} + 2\mathrm{H}_2\mathrm{O}$ & (7)\\
$\mathrm{P} + 3\mathrm{H}_2\mathrm{O} + 3e^{-} \rightleftharpoons \mathrm{PH}_3 + 3\mathrm{OH}^{-}$ & (8)\\
$\mathrm{SO}_4^{2-} + \mathrm{H}_2\mathrm{O} + 2e^{-} \rightleftharpoons \mathrm{SO}_3^{2-} + 2\mathrm{OH}^{-}$ & (9)\\
$\mathrm{PO}_4^{3-} + 2\mathrm{H}_2\mathrm{O} + 2e^{-} \rightleftharpoons \mathrm{HPO}_3^{2-} + 3\mathrm{OH}^{-}$ & (10)\\
$\mathrm{HPO}_3^{2-} + 2\mathrm{H}_2\mathrm{O} + 3e^{-} \rightleftharpoons \mathrm{P} + 5\mathrm{OH}^{-}$ & (11)\\
$\mathrm{SO}_4^{2-} + 4\mathrm{H}^{+} + 2e^{-} \rightleftharpoons \mathrm{H}_2\mathrm{SO}_3 + \mathrm{H}_2\mathrm{O}$ & (12)\\
$\mathrm{H}_2\mathrm{SO}_3 + 4\mathrm{H}^{+} + 4e^{-} \rightleftharpoons \mathrm{S} + 3\mathrm{H}_2\mathrm{O}$ & (13)\\
$\mathrm{Cl}_2 + 2e^{-} \rightleftharpoons \mathrm{Cl}^{-}$ & (14)\\
\end{tabular}
\caption{Table showing all the half-reactions present during the electrodeposition of Ni-P as outlined by Daly \& Barry\cite{NiP-Table}.}
\label{Table NiP Half Reaction}
\end{table}

Given the time-sensitive and kinetically driven nature of the EHD-RP process, the direct mechanism of depositing solid P is far more promising. Thus, H\textsubscript{3}PO\textsubscript{2} was selected for these experiments as it can be directly reduced to elemental phosphorus in an acidic environment.  
\end{document}